\documentclass[10pt, aps, prl, onecolumn]{revtex4-2}  % More conventional
\usepackage{ulem}
\usepackage{amsmath, amssymb}
\usepackage{xcolor}
\usepackage{overpic}
\usepackage{braket, upgreek}
\usepackage{BOONDOX-cal}
\usepackage{enumitem}
\usepackage{graphicx}
\usepackage{caption} % Preferred over subfigure
\usepackage{float}
\usepackage{orcidlink}
\usepackage{tikz}
\usepackage{subcaption}
\usepackage{subcaption}
\usepackage{xcolor}
\usepackage{xcolor}
\usepackage{xcolor}
\usepackage{lipsum}
\setlist[enumerate]{label=\arabic*., leftmargin=*}
\usepackage[utf8]{inputenc}

\DeclareMathAlphabet{\boondoxmathcal}{U}{BOONDOX-cal}{m}{n}

\DeclareMathAlphabet{\boondoxmathbcal}{U}{BOONDOX-cal}{b}{n}
\DeclareMathAlphabet{\mathcal}{OMS}{cmsy}{m}{n}

\newcommand{\e}{\boondoxmathcal{e}}
\newcommand{\p}{\boondoxmathcal{p}}

\makeatother

\date{\today}
\begin{document}

\title{Characterization of Polariton Dynamics in a Multimode Cavity (II): Coherent-Incoherent Transition Driven by Photon Loss}

\author{Md Qutubuddin\,\textsuperscript{1,\orcidlink{0000-0002-1294-5154}}}
\thanks{These authors contributed equally to this work.}
%%%%%%%%%%%%%%%%%%%%%%%%%%%%%%%%%%%%%%%%%%%%%%%%%%%%%%%%%%%%%%%%%%%%%%%%%%%%%%%%%%%%
\author{Ilia Tutunnikov\,\textsuperscript{2,\orcidlink{0000-0002-8291-7335}}}
\thanks{These authors contributed equally to this work.}
%%%%%%%%%%%%%%%%%%%%%%%%%%%%%%%%%%%%%%%%%%%%%%%%%%%%%%%%%%%%%%%%%%%%%%%%%%%%%%%%%%%
\author{Jianshu Cao\,\textsuperscript{2,\orcidlink{0000-0001-7616-7809}}}
\email{jianshu@mit.edu}
\affiliation{\textsuperscript{1} Beijing Computational Science Research Center, Beijing 100193, China}
\affiliation{\textsuperscript{2} Department of Chemistry, Massachusetts Institute of Technology, Cambridge,
Massachusetts 02139, USA}

\begin{abstract}
Motivated by the recent advances in optical imaging and tracking of wave-packet propagation in optical cavities, we systematically explore the non-Hermitian polariton dynamics within a decay-tunable multimode cavity model.  The complex eigen-spectrum of the model Hamiltonian allows us to predict the incoherent-coherent transition induced by photon losses, which defines an exceptional point at resonance and evolves analytically as the wavevector shifts off-resonantly.  The resulting dispersion relation, group velocity, and relaxation rate exhibit striking signatures, such as curve crossing, level repulsion, turnover, bifurcation, and coalescence, as the decay rate crosses the critical transition or the wavevector crosses the resonance. The spectral characterization leads to surprising features in the non-Hermitian wave packet dynamics: (i) maximal population relaxation rate at the critical transition; (ii) reversed propagation in the center-of-mass motion; (iii) ballistic-to-diffusion transition; (iv) contraction in the displacement and width of the polariton wave-packet. These dynamical features have complementary symmetry between the upper-polariton (UP) branch and lower-polariton (LP) branch in the two-dimensional phase diagram spanned by the photon decay rate and wavevector.  Thus, the combination of complex spectral characterization and non-Hermitian wave packet propagation establishes the photon decay rate as a powerful control parameter for polariton transport, reveals the underlying symmetry in lossy cavities, and presents a starting point to incorporate other dissipative mechanisms.
%%%%%%%%%%%%%%%%%%%%%%%%%%%%%%%%%%%%%%%%%%%%%%%%%%%%%%%%%%%%%%%%%%%%%%%%%%%%%%%%%%%%%%%%%%%%%%%%%%%%%%%%%
% We investigate the non-Hermitian dynamics of strongly coupled cavity-molecule systems through a theoretical framework based on a decay-tunable multimode polariton Hamiltonian. Our analysis reveals how controlled cavity dissipation quantitatively modifies both the energy spectrum (including its momentum derivative) and transport properties for both upper and lower polariton branches. 
% %%%%%%%%%%%%%%%%%%%%%%%%%%%%%%%%%%%%%%%%%%%%%%%%%%%%%%%%%%%%%%%%%%%%%%%%%%%%%%%%%%%%%%%
% Using exact diagonalization of the non-Hermitian system, we first identify critical points in the parameter space and characterize wavevector-dependent spectral features as functions of the cavity loss rate. The modified energy landscape leads to three well-defined dynamical regimes: (i) coherent propagation at weak dissipation, (ii) critical behavior at intermediate loss, and (iii) incoherent dynamics in the strong dissipation limit.
% %%%%%%%%%%%%%%%%%%%%%%%%%%%%%%%%%%%%%%%%%%%%%%%%%%%%%%%%%%%%%%%%%%%%%%%%%%%%%%%%%%%%%%%%%%%%%
% We provide a complete characterization of the coherent-to-incoherent transition and its impact on wavepacket evolution through detailed analytical and numerical studies of center-of-mass motion and mean-squared displacement. These results establish cavity loss rate as a powerful control parameter for polariton transport in organic quantum materials.
\end{abstract}
\maketitle
\section{Introduction}
The study of non-Hermitian quantum systems has garnered significant attention due to their unique physical properties and potential applications in quantum optics, photonics, and condensed matter physics. In particular, the interplay between gain, loss, and coherent coupling in such systems leads to exotic phenomena, including exceptional points \cite{rvwali2019, friederike2025sa}, parity-time (PT) \cite{rvwganainy2018} symmetry breaking~\cite{jianming2024ncom}, and non-trivial topological states \cite{hu2023prb, frank2022nanop, rui2021sca}. In this context, the proposed multi-mode non-Hermitian Tavis-Cummings (MNTC) model extends the Tavis-Cummings model~\cite{tavis1968phyrev} to incorporate complex photonic environments, offering a lossy theoretical framework for investigating polariton dynamics. 

Polaritons, as hybrid light–matter quasiparticles \cite{rvwsandik2024}, offer a natural platform for realizing and studying non-Hermitian effects in strongly coupled photonic systems. In optical microcavities with quality factors \( Q \sim 20{-}10^3 \) \cite{Hobson2002apl}, corresponding to cavity loss rates in the range of \( 0.1{-}0.001 \, \text{eV} \) for a resonance energy \( \omega_C = 2 \, \text{eV} \), polaritons emerge from strong exciton–photon coupling and exhibit robust quantum coherence. In contrast, plasmonic polaritons(e.g., plexcitons), which arise from coupling between excitons and plasmonic cavities, typically exhibit lower quality factors \( Q \sim 1{-}10 \), corresponding to higher cavity loss rates of \( 2{-}0.2 \, \text{eV} \) \cite{baranov2018acs, chikka2016nat}. While such systems enable extreme light confinement and strong hybridization, they are also subject to significant losses and decoherence~\cite{Torma2015rprt, antosiewicz2014acs}. 
%%%%%%%%%%%%%%%%%%%%%%%%%%%%%%%%%%%%%%%%%%%%%%%%%%%%%%%%%%%%%%%%%%%%%%%%%%%%%
In the multi-mode non-Hermitian Tavis-Cummings (MNTC) framework, non-Hermitian effects play a key role in capturing how dissipation and mode coupling affects the formation, relaxation, and transport of polaritons.\cite{tichauer2021jcp}.
%%%%%%%%%%%%%%%%%%%%%%%%%%%%%%%%%%%%%%%%%%%%%%%%%%%%%%%%%%%%%%%%%%%%%%%%%%%%%%%%%%%%%%%%%%%%%%%%%%%%%

In this article, we present a comprehensive analysis of microcavity polariton and plexciton wave packets within the MNTC frame. We explore the full spectral behavior across photon loss rates, identifying the critical transition between the coherent and incoherent regimes for both the UP and LP branches. Resonant and off-resonant regimes reveal distinct dynamical features: one branch shows non-monotonic population relaxation, reversal of wave packet (WP) direction, and ballistic to diffusive transition, while the other shows monotonic behavior in population relaxation, unidirectional WP propagation~\cite{tutunnikov2023acs}, and WP contraction \cite{pandaya2022sciadv, chng2025nano}. The resonance behavior and the critical transition define a \( 2D \) phase diagram with \(4\) complementary quadratures (see Fig.~\ref{fig:cartoon_1}).
%%%%%%%%%%%%%%%%%%%%%%%%%%%%%%%%%%%%%%%%%%%%%%%%%%%%%%%%%%%%%%%%%%%%%%%%%%%%%%%%%%%%%%%%%%%%%%%%%%%%%%%%%%
\section{Multimode non-Hermitian Tavis-Cumming model} \label{sec:mntc_hamiltonian}
We consider a one-dimensional array of length $L$ containing $N$ quantum emitters, which model fundamental excitation transitions in atoms, molecules, qubits, or semiconductors. To model the interaction between molecular excitations and a multimode leaky cavity, we describe the exciton subsystem using the Bloch states $\propto\exp[i2\pi kn/N]$ (where $k$ is the
wave number, $n$ is the lattice index). This basis couples each cavity mode $k$  to a corresponding exciton mode with the same wave number, effectively partitioning the system into independent $k$-components. The total non-Hermitian Hamiltonian written as $\hat{H} = \sum_{k} \hat{H}_{k}$, where $\hat{H}_{k}$ governs the interaction between $k_{th}$ cavity mode and $k_{th}$ exciton mode, giving
\begin{equation}
\hat{H}_{k}=\begin{bmatrix}\omega_{M} & g\\
g & \omega_{k}-i\gamma
\end{bmatrix},\label{eq:2x2-H}
\end{equation}
where $\gamma\geq0$ is the photon loss rate, $\omega_M$ is the molecular energy, and $\omega_k$ is the $k_{th}$ mode of photon energy. The photon dispersion $\omega_{k}=\sqrt{\left(\frac{2\pi k}{N}\right)^{2}+\omega_{C}^{2}}$, where $\omega_{C}$ is photon confinement energy depending on the geometry of the microcavity,  and $N$ is the number of quantum emitters~\cite{tutunikov2025} (which is set to equal to the number of modes). 

The eigenvalues and eigenvectors of the MNTC Hamiltonian are
\begin{equation} 
\varepsilon_{k_\pm} = \frac{ \omega_M + \omega_k - i \gamma \pm \Delta_k }{2}, \label{eq:eigenenergies}
\end{equation}
\begin{equation}
\ket{v_{k_{\pm}}}=\begin{cases}
\e_{k_{+}}\ket{b_{k}}+\p_{k_{+}}\ket{a_{k}}, & + \\
\e_{k_{-}}\ket{b_{k}}+\p_{k_{-}}\ket{a_{k}}, & -
\end{cases},\label{eq:eigenstates}
\end{equation}
%%%%%%%%%%%%%%%%%%%%%%%%%%%%%%%%%%%%%%%%%%%%%%%%%%%%%%%%%%%%%%%%%%%%%%%%%%%%%%%%%%%%%%%%%%
where $\pm$ denotes the upper polariton (UP) and lower polariton (LP) branches. Here, the effective Hopfield coefficients are given by
$\e_{k_{\pm}}  =\frac{\varepsilon_{k_{\mp}}-\omega_{M}}{\sqrt{(\varepsilon_{k_{\mp}}-\omega_{M})^{2}+g^{2}}},
\p_{k_{\pm}}  =\frac{g}{\sqrt{(\varepsilon_{k_{\mp}}-\omega_{M})^{2}+g^{2}}}$, and $\Delta_k$ represents the complex-valued vacuum Rabi splitting (VRS), defined as:
%%%%%%%%%%%%%%%%%%%%%%%%%%%%%%%%%%%%%%%%%%%%%%%%%%%%%%%%%%%%%%%%%%%%%%%%%%%%%%%
\begin{equation} \label{rabi_frequency}
\Delta_k \equiv \varepsilon_{k_+}-\varepsilon_{k_-} = \sqrt{(\delta_k + i \gamma)^2 + 4g^2}, \end{equation}
%%%%%%%%%%%%%%%%%%%%%%%%%%%%%%%%%%%%%%%%%%%%%%%%%%%%%%%%%%%%%%%%%%%%%%%%%%%%%%%%
where $\delta_k = \omega_M - \omega_k$ is the detuning. The VRS spectrum exhibits a discontinuous transition by varying $\delta_k$ and $\gamma$, corresponding to phase jumps between quadrants in the complex plane defined by  \( \text{Re}[\![\Delta^2_k]\!] \ \text{and} \ \text{Im}[\![\Delta^2_k]\!] \). For the VRS condition \(\Delta_k = 0\), the photon loss rate reaches a critical point,
\( \gamma_C = 2g = 0.6. \) At resonance, where the detuning vanishes \(\delta_k = 0\), the wavevector is \( k_r = 0.92. \) These transitions explain the unusual behaviors observed in relaxation, dispersion, and group velocity  $v_g$, as will be analyzed in subsequent paragraphs and detailed in Supplementary Information (SI). %\ref{si:phase_analysis}.

The real parts of the eigenvalues in Eq.~\eqref{eq:eigenenergies} define the dispersion curves of the UP and LP branches;
%%%%%%%%%%%%%%%%%%%%%%%%%%%%%%%%%%%%%%%%%%%%%%%%%%%%%%%%%%%%%%%%
\begin{eqnarray}
\text{Re}[\varepsilon_k{}_\pm] = \frac{\omega_M + \omega_k \pm \text{Re}\Delta_k}{2}, \label{eq:dispersion}
\end{eqnarray}
and imaginary parts of Eq.~\eqref{eq:eigenenergies} 
define the characteristic relaxation rates associated with the UP and LP branches:
%%%%%%%%%%%%%%%%%%%%%%%%%%%%%%%%%%%%%%%%%%%%%%%%%%%%%%%%%%%%%%%%%%%%%%%%%%%%%%%%
\begin{equation} 
\gamma_{k_\pm}  =  \gamma \mp \text{Im}[\Delta_k]. \label{eq:decay_rate}
\end{equation}
\begin{figure}[h]
    \centering
    % \begin{subfigure}{0.51\textwidth} % Reduced width to 0.45
    \includegraphics[width=\linewidth]{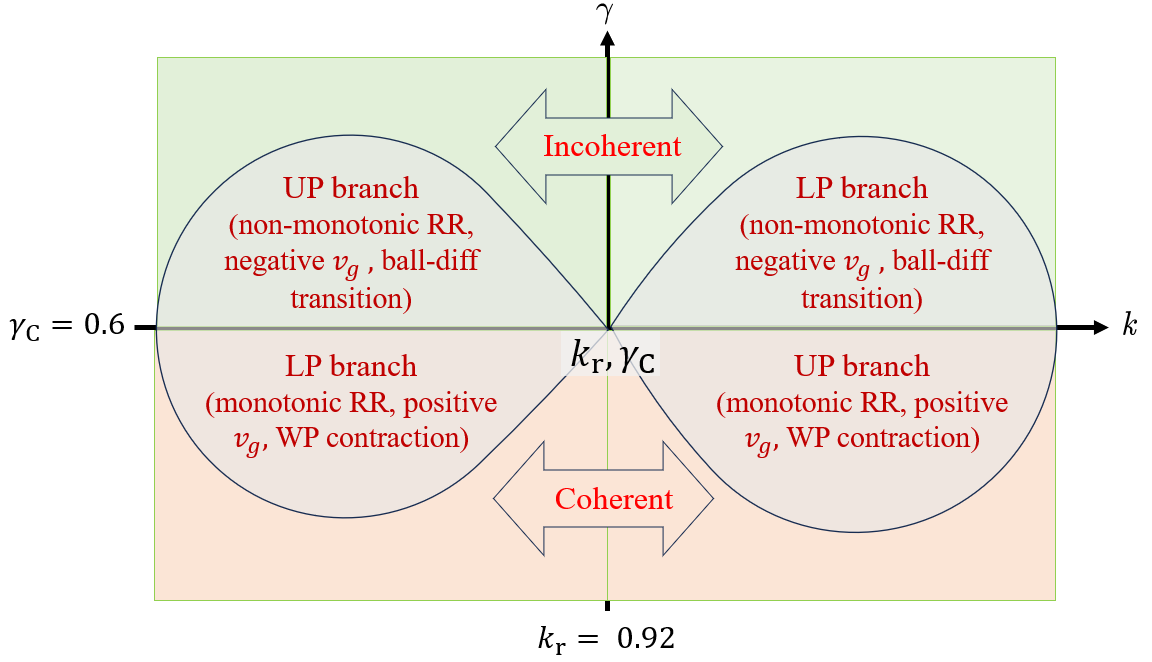} % Ensure image fits within subfigure
    \caption{Schematic illustration of the wave vector and photon loss rate, highlighting the resonance wave vector and critical photon loss rate at the origin. RR - relaxation rate and WP - wave packet.}
    \label{fig:cartoon_1}
    % \end{subfigure}
\end{figure}
The full spectrum of the non-Hermitian Hamiltonian and its dynamical properties are summarized in Fig.~\ref{fig:cartoon_1}. In this schematic representation, the origin \((k_r, \gamma_C)\) denotes the resonance point and the critical point, respectively. The \(x\)-axis represents the wave vector \(k\), while the \(y\)-axis represents the photon loss rate \(\gamma\). The phase diagram in Fig.~\ref{fig:cartoon_1} is supported by the eigenvalue analysis in Fig.~\ref{fig:disp_grpvel}, which quantitatively characterizes the UP/LP branches. The Hopfield coefficients (represented by the blue gradient, which quantifies the light-matter hybridization strength), wave vector, and photon loss rates collectively determine the dispersion, group velocity, and relaxation properties of polariton branches, as we analyze in detail below.

%%%%%%%%%%%%%%%%%%%%%%%%%%%%%%%%%%%%%%%%%%%%%%%%%%%%%%%%%%%%%%%%%%%%%%%%%%%%%%%%%%%%%%%%%%%%%%%%%%%%%%%
\subsection{Dispersion}\label{subs:dispersion}
\begin{figure*}
    \centering
    \begin{subfigure}{0.55\textwidth} % Reduced width to 0.45
        \centering
        \includegraphics[width=\linewidth]{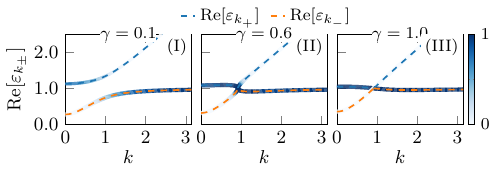} % Ensure image fits within subfigure
        \caption{}
        \label{fig:dispersion_k}
    \end{subfigure}
    \begin{subfigure}{0.55\textwidth} % Reduced width to 0.45
        \centering
        \includegraphics[width=\linewidth]{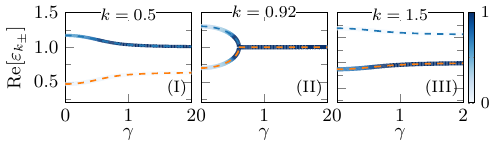} % Ensure image fits within subfigure
        \caption{}
        \label{fig:dispersion_g}
    \end{subfigure}
     \begin{subfigure}{0.55\textwidth} % Reduced width to 0.45
        \centering
        \includegraphics[width=\linewidth]{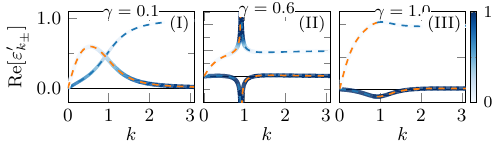} % Ensure image fits within subfigure
        \caption{}
        \label{fig:vg_k}
    \end{subfigure}
    % \hfill
    \begin{subfigure}{0.55\textwidth} % Reduced width to 0.45
        \centering
        \includegraphics[width=\linewidth]{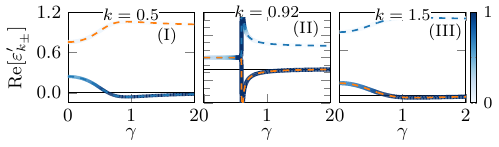} % Ensure image fits within subfigure
        \caption{}
        \label{fig:vg_g}
    \end{subfigure} 
    %%%%%%%%%%%%%%%%%%%%%%%%%%%%%%%%%%%%%%%%%%%%%%%%%%%%%%%%%%%%%%%%%%%%%%%%%%%%
    \begin{subfigure}{0.57\textwidth} % Reduced width to 0.45
        \centering
        \includegraphics[width=\linewidth]{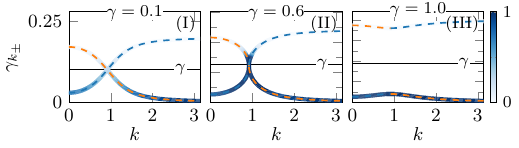} % Ensure image fits within subfigure
        \caption{}
        \label{fig:decay_rate_k}
    \end{subfigure}
    % \hfill
    \begin{subfigure}{0.57\textwidth} % Reduced width to 0.45
        \centering
        \includegraphics[width=\linewidth]{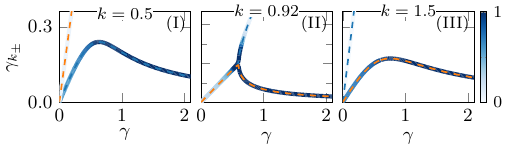} % Ensure image fits within subfigure
        \caption{}
        \label{fig:decay_rate_g}
    \end{subfigure} 
    %%%%%%%%%%%%%%%%%%%%%%%%%%%%%%%%%%%%%%%%%%%%%%%%%%%%%%%%%%%%%%%%%%%
    \caption{(a, b): Polariton dispersion curve $\text{Re}\varepsilon_{k_\pm}$ of the UP/LP as a function of the wave vector and photon loss rate, respectively. (c, d): Group velocity, Re$[\varepsilon'_k{}_\pm]$ of the UP/LP branches. (e, f): Relaxation rate in Eq.~(\ref{eq:decay_rate}) of the UP/ LP branches. $\text{{Parameters:\;}}\omega_{M}=1,\,\omega_{0}=0.4,\,g=0.3$.}
    \label{fig:disp_grpvel}
\end{figure*}

Fig.~\ref{fig:dispersion_k} is the dispersion curve, \(\text{Re}[\varepsilon_{k\pm}]\) for three photon loss rates: \(\gamma = 0.1\) (coherent regime), \(0.6\) (critical regime), and \(1.0\) (incoherent regime). These representative values will be used throughout our subsequent analysis.
Fig. ~\ref{fig:dispersion_k}(I) shows the dispersion relation from Eq. (\ref{eq:dispersion}) in the coherent regime (\(\gamma < \gamma_C \)), where the UP and LP branches exhibit avoided crossing due to level repulsion. In contrast, Fig.~\ref{fig:dispersion_k}(II) at the critical point ( \(\gamma = \gamma_C\) ) exhibits crossing at resonance as a result of level attraction. Fig.~\ref{fig:dispersion_k}(III), corresponding to the incoherent regime \( (\gamma > \gamma_C) \), shows a flattened crossing at resonance reflecting a suppressed dispersion, which is a unique signature of the non-Hermitian Hamiltonian. 

We now analyze the real part of the eigenvalues, \( \text{Re}[\varepsilon_{k\pm}] \), as a function of the photon loss rate \( \gamma \), for three representative wave vector values, as shown in Fig.~\ref{fig:dispersion_g}. The off-resonant case \( k = 0.5 < k_r \) in Fig.~\ref{fig:dispersion_g}(I), where \( \text{Re}[\varepsilon_{k\pm}] \) are plotted as functions of the photon loss rate \( \gamma \). The UP branch maintains a high energy, while the LP branch remains at lower energy, with both exhibiting weak dependence on $\gamma$.
%%%%%%%%%%%%%%%%%%%%%%%%%%%%%%%%%%%%%%%%%%%%%%%%%%%%%%%%%%%%%%%%%%%%%%%%%%%%%%%%%%%%%%%%%%%%%%%%%%%
In contrast, for the resonant \( k = k_r \) in Fig.~\ref{fig:dispersion_g}(II), two branches of the coherent regime coalesce at the critical point $\gamma_C$ into a single line in the incoherent regime. The separation between these branches follows a parabolic function as a function of $\gamma_C - \gamma$, as detailed in SI~IB. The reduction of the Rabi splitting and its coalescence at the transition point is consistent with observations from our exception point analysis of disordered cavities \cite{engelhardt2023prl}. In this context, the photon loss rate \( \gamma \) in the present study plays a role analogous to the disorder strength considered in our previous investigations~\cite{engelhardt2022prb}. As $k = 1.5 > k_r$ in Fig.~\ref{fig:dispersion_g}(III), the UP/LP branches exhibit a weak dependence on $\gamma$, analogous to their behavior for $k < k_r$. However, the energy ordering of the UP/LP branches inverted relative to the case of $k < k_r$.
% \setlength{\abovecaptionskip}{0pt}
%%%%%%%%%%%%%%%%%%%%%%%%%%%%%%%%%%%%%%%%%%%%%%%%%%%%%%%%%%%%%%%%%%%%%%%%%%%%%%%%%%%%%%%%%%%%%%%%%%%
\subsection{Group velocity}\label{subs:grp_velocity}
\textcolor{black}{Having analyzed the eigenenergies as functions of the wave vector \( k \) and photon loss rate \( \gamma \)}, we now examine the corresponding group velocities, defined as \( v_g = \text{Re} [\varepsilon'_{k_{\pm}}] \), which are plotted in Fig.~\ref{fig:vg_k} as a functions of \(k\). These group velocities determine the propagation characteristics of polaritonic wave packet (WP). In the coherent regime \( (\gamma < \gamma_C) \), Fig.~\ref{fig:vg_k}(I) shows the distinct behavior of the LP and UP branches: (i) The group velocity of the LP peaks at \( k < k_r \), where the dispersion is steepest due to strong photon hybridization, whereas the group velocity of the UP increases monotonically with \( k \). (ii) At $k=k_r$, where the hybridization is maximal, the group velocities of the LP and UP branches cross as their dispersion begins to flatten. (iii) For $k > k_r$, the group velocity of the LP approaches zero due to increasing excitonic character, while \( v_g\) of the UP saturates to the photon group velocity. In the incoherent regime \( \gamma \ge \gamma_C \), maximal hybridization occurs at the resonance point \( k = k_r \), where \( v_g \) of the LP branch reaches its peak. Notably, $v_g$ become negative for $\gamma > \gamma_C$, in Figs.~\ref{fig:vg_k}(II) and \ref{fig:vg_k}(III). This sign reversal indicates that the corresponding wave packets propagate in the opposite direction to their initial momentum, revealing anomalous transport behavior characteristic of the system's non-Hermitian nature. 

Next, we analyze the derivative of the group velocity, \( v_g = \text{Re}[\partial_k \varepsilon_{k\pm}] \), for the UP and LP branches as a function of the photon loss rate \( \gamma \), as shown in Fig.~\ref{fig:vg_g} for three representative values of \( k \). Fig.~\ref{fig:vg_g}(I) shows the $\gamma$ dependence of group velocity for the UP/LP branches in the off-resonance case $k = 0.5 < k_r$: $v_g$ of the UP decreases with increasing $\gamma$, and it turns negative at the critical point $(\gamma = \gamma_C)$ and remains negative while approaching zero for large $\gamma$. Conversely, $v_g$ of the LP increases with $\gamma$, reaches maxima at the critical point, and subsequently saturates to a positive value for large $\gamma$. At resonance $k = k_r$, Fig.~\ref{fig:vg_g}(II) shows the bifurcation at a critical point, where $v_g$ of the LP becomes negative, while \( v_g \) of the UP turns positive. Beyond the critical point, $v_g$ of the LP approaches zero, whereas $v_g$ of the UP approaches to finite positive value, with increasing $\gamma$. As $k = 1.5 > k_r$, Fig.~\ref{fig:vg_g}(III) exhibits a behavior similar to that observed for $k<k_r$. However, $v_g$ of these branches reversed: $v_g$ of the UP remains positive, while \( v_g\) of the LP undergoes a sign change from positive to negative at the critical point $\gamma_C$. 
%%%%%%%%%%%%%%%%%%%%%%%%%%%%%%%%%%%%%%%%%%%%%%%%%%%%%%%%%%%%%%%%%%%%%%%%%%%%%%%
\subsection{Relaxation rate}\label{subs:relaxation_rate}
We now analyze the imaginary part of the eigenenergies of the MNTC Hamiltonian, which governs the relaxation rates of the UP and LP branches, as described in Eq.~\eqref{eq:decay_rate}. We first analyze its dependence on the wave vector \(k\) (Fig.~\ref{fig:decay_rate_k}), followed by its dependence on the photon loss rate \(\gamma\) (Fig.~\ref{fig:decay_rate_g}).

In Fig.~\ref{fig:decay_rate_k}(I) for the coherent regime \( ( \gamma < \gamma_C ) \), the LP relaxation rate decreases monotonically, while the UP increases monotonically with increasing $k$. Both rates exhibit symmetry around the resonance point, where they cross. At large $k$, the LP relaxation rate approaches zero, whereas the UP relaxation rate saturates to a finite value. In Fig.~\ref{fig:decay_rate_k}(II), at the critical point (\( \gamma = \gamma_C\)), the trends of the LP and UP relaxation rates are similar to those in Fig.~\ref{fig:decay_rate_k}(I) but with enhanced magnitude and sharp transition near $k_r$. In Fig.~\ref{fig:decay_rate_k}(III), within the incoherent regime (\(\gamma > \gamma_C\)), a sudden jump in the relaxation rates of the UP/LP  branches is observed at resonance. This discontinuity manifests as a gap in the relaxation rate, signaling the transition from a coherent to an incoherent regime, where the system no longer supports hybridized light-matter modes. This intriguing $k$-dependence behavior arises from phase variation as predicted in SI~(II.1).
%\ref{si:phase_analysis}.
 
%%%%%%%%%%%%%%%%%%%%%%%%%%%%%%%%%%%%%%%%%%%%%%%%%%%%%%
Next, we analyze the dependence of the relaxation rate on the photon loss rate \(\gamma\). In the off-resonant regime (\(k < k_r\)), Fig.~\ref{fig:decay_rate_g}(I) reveals distinct behaviors for the UP and LP branches: the relaxation rate of the UP branch exhibits a pronounced peak at the critical point, reflecting the critical enhancement of photonic dissipation, whereas the relaxation rate of the LP branch increases approximately linearly with \(\gamma\) (see details in SI~(II.2)).
%%%%%%%%%%%%%%%%%%%%%%%%%%%%%%%%%%%%%%%%%%%%%%%%%%%%%%%%%%%%%%%%%%%%%%%%%%%%%%%%%%%%%%%%%%%%%%%%%%%%%%
At resonance $k=k_r$, Fig.~\ref{fig:decay_rate_g}(II) shows the bifurcation at the critical point from a single line in the coherent regime to two distinct branches. Near this transition, the relaxation rates trace out a characteristic parabolic curve, signaling the onset of mode splitting.
This bifurcation was also observed in our previous study of disorder cavities and is associated with the emergence of a critical point underlying a non-Hermitian Hamiltonian. 
%%%%%%%%%%%%%%%%%%%%%%%%%%%%%%%%%%%%%%%%%%%%%%%%%%%%%%%%%%%%%%%%%%%%%%%%%%%%%%%%%%%%%%%%%%%%%%%%%%%%%%%%%%
Beyond resonance $k>k_r$ as shown in Fig.~\ref{fig:decay_rate_g}(III), while the $\gamma$ dependence of relaxation rate exhibits similar characteristics to $k<k_r$, the role of the UP/LP branches is reversed.
\begin{figure}[h]
    \centering
  % First row
    \begin{subfigure}{0.4\linewidth}
        \includegraphics[width=\linewidth]{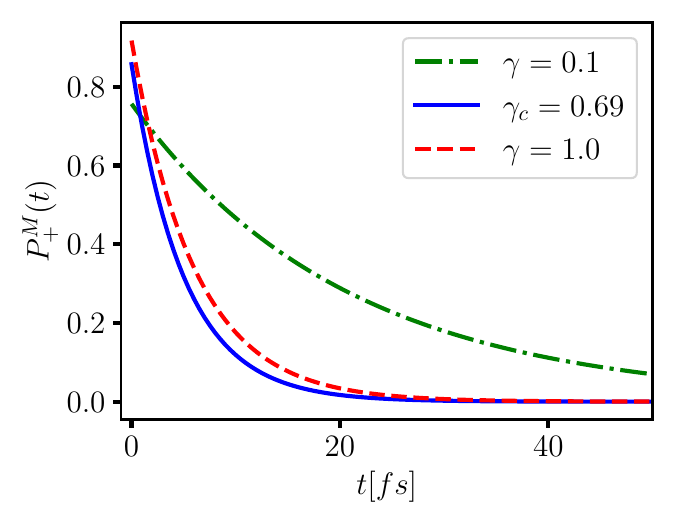}
        \caption{}
    \label{fig:pop_up_exc}
    \end{subfigure}
    % \hfill
    \begin{subfigure}{0.4\linewidth}
        \includegraphics[width=\linewidth]{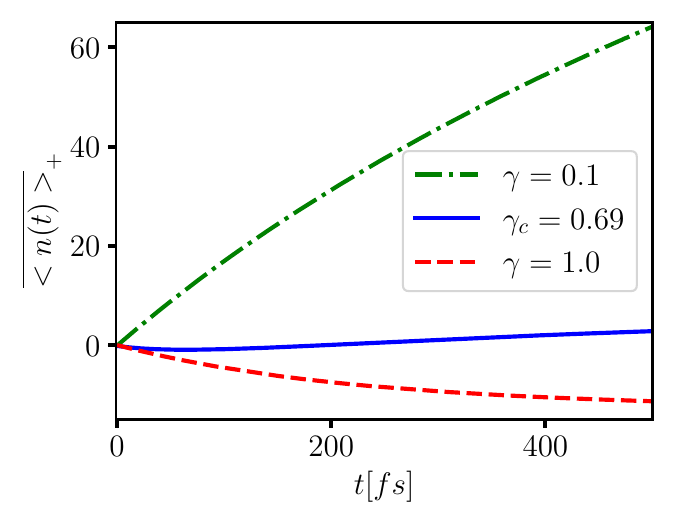}
        \caption{}
        \label{fig:cm_up}
    \end{subfigure}
    % \vspace{0.5cm} 
    % Second row
    \begin{subfigure}{0.4\linewidth}
        \includegraphics[width=\linewidth]{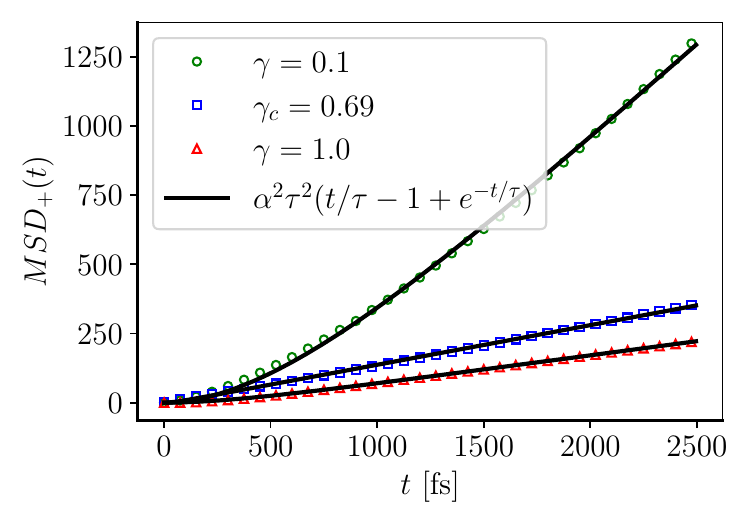}
        \caption{}
        \label{fig:msd_up}
    \end{subfigure}
    % \hfill
    \begin{subfigure}{0.38\linewidth}
        \includegraphics[width=\linewidth]{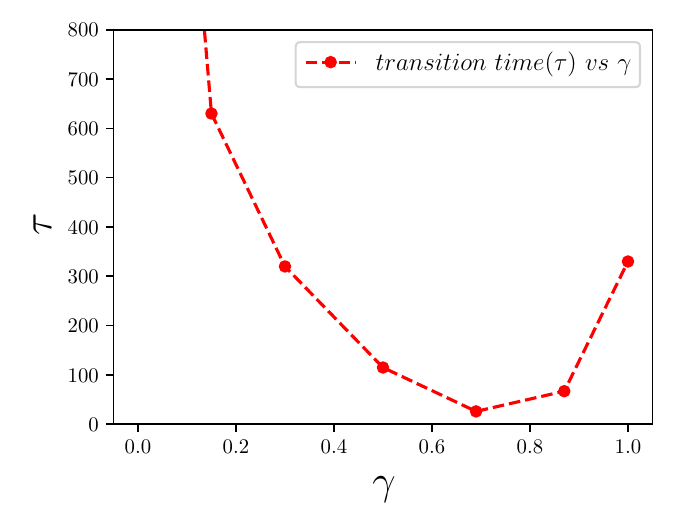}
        \caption{} % Starred version suppresses numbering
        \label{fig:transition_gamma}
    \end{subfigure}
    \caption{(a) The population dynamics of the UP branch are shown for three photon loss rates: \(\gamma = 0.1\), \(0.67\), and \(1.0\). (b) The normalized center-of-mass (CM) position propagation of the UP branch is shown for the same set of photon loss rates. (c) The mean squared displacement (MSD) of the UP wave packet dynamics with a fitting function $f(t)$ as evidenced by the bold fitted curve, which follows the expected theoretical expression. (d) The ballistic transition of the UP parameters illustrated against $\gamma$. Parameters: $\omega_M = 1.0, \omega_C = 0.4, g = 0.3, p = 0.5$}
    \label{fig:upper_polariton}
\end{figure}
%%%%%%%%%%%%%%%%%%%%%%%%%%%%%%%%%%%%%%%%%%%%%%%%%%%%%%%%%%%%
\section{wave packet dynamics}
\subsection{Exciton population relaxation}
Here we present the dynamics at three key $(\gamma, k)$ combinations identified in the earlier section. To understand the collective behavior of the UP and LP branches, we derive analytical expressions for the population dynamics of the wave packets. The wave function of the exciton subsystem for the UP and LP branches in $k$-space is given by
\begin{equation}
\psi^{M}_{k_\pm}(t)=\psi_{k}(0) \e_{k_{\pm}}e^{-i\varepsilon_k{}_\pm t},\label{eq:molwave-function}
\end{equation}
where $\psi_{k}(0) = \sqrt{w/ \sqrt{\pi}}\exp\left[-w^{2}/2(k-p)^{2}\right]$ is the initial amplitude of the exciton mode $k$.
% \begin{equation}
% \psi_{k}(0)=\sqrt{\frac{w}{\sqrt{\pi}}}\exp\left[-\frac{w^{2}}{2}(k-p)^{2}\right].\label{eq:=00005Cpsi_k(0')}
% \end{equation}
%\section{Populations dynamics}
The population distribution of $k$-th exciton subsystem of the UP/LP follows, $P^{M}_{k_{\pm}}(t)= |\psi_{k_\pm}^{M }(t)|^2$, and the exciton population is given by
\begin{equation}
P^{M}_{\pm}(t) = \int^{\pi}_{-\pi} P^{M}_{k_\pm} (t) dk =\int_{-\pi}^{\pi} P_{k}(0) |\e_{k_{\pm}}|^{2}e^{-\gamma_{k_{\pm}} t} dk. \label{eq:total_exc_pop}
\end{equation}
%%%%%%%%%%%%%%%%%%%%%%%%%%%%%%%%%%%%%%%%%%%%%%%%%%%%%%%%%%%%%
where $P_{k}(0) = | \psi_{k}(0)|^{2}$. 
%%%%%%%%%%%%%%%%%%%%%%%%%%%%%%%%%%%%%%%%%%%%%%%%%%%%%%%%%%%%%%%%%%%%%%%%%%%%%%%%%%%%%%%%%%%%%
The propagation of the total exciton population (Eq.~\ref{eq:total_exc_pop}) in the UP branch is shown in Fig.~\ref{fig:pop_up_exc}. The relaxation rate of the exciton population increases with the photon loss rate \( \gamma \), reaching a maximum at the critical point \( \gamma_C \), and decreases beyond this value. This turnover in the relaxation rate of the total excitonic population of the UP branch reveals an intriguing experimental implication: even under strong decoherence conditions (\( \gamma \gg \gamma_C \)), the population can persist for remarkably long times. The detailed analysis of population dynamics is discussed in SI~(III.1). %~\ref{si:population_dynamics}.
%%%%%%%%%%%%%%%%%%%%%%%%%%%%%%%%%%%%%%%%%%%%%%%%%%%%%%%%%%%%%%%%%%%%%%%%%%%%%%%
\begin{figure}[h]
    \centering
  % First row
    \begin{subfigure}{0.41\linewidth}
      \includegraphics[width=\linewidth]{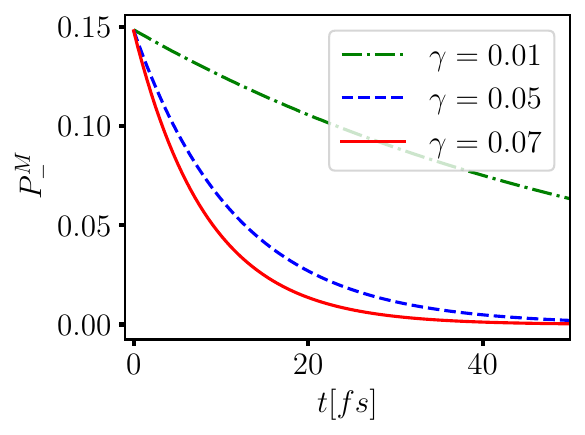}
        \caption{}
        \label{fig:pop_lp_exc}
    \end{subfigure}
    % \hfill
    \begin{subfigure}{0.41\linewidth}
        \includegraphics[width=\linewidth]{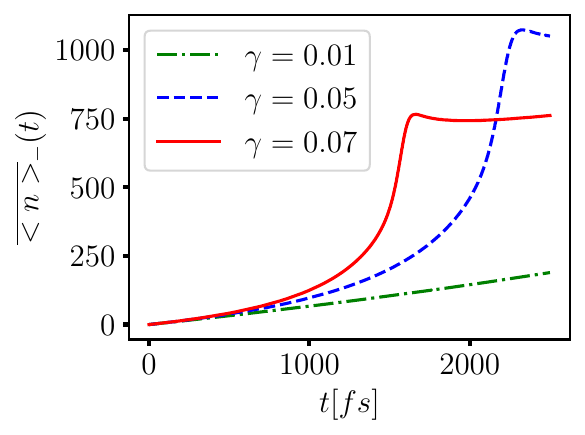}
        \caption{}
        \label{fig:cm_lp}
    \end{subfigure}
    % \vspace{0.5cm} 
    % Second row
    \begin{subfigure}{0.41\linewidth}
        \includegraphics[width=\linewidth]{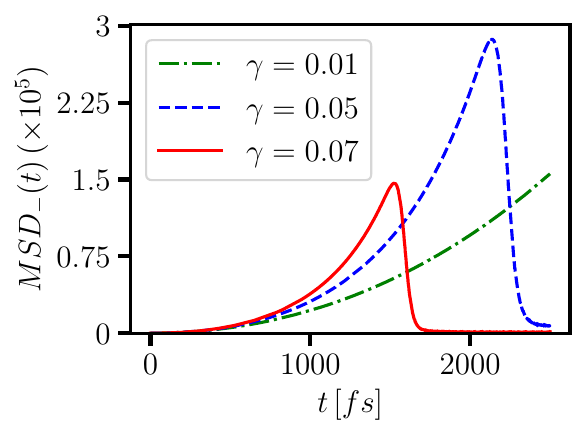}
        \caption{}
        \label{fig:msd_lp}
    \end{subfigure}
    % \hfill
    \begin{subfigure}{0.41\linewidth}
        \includegraphics[width=\linewidth]{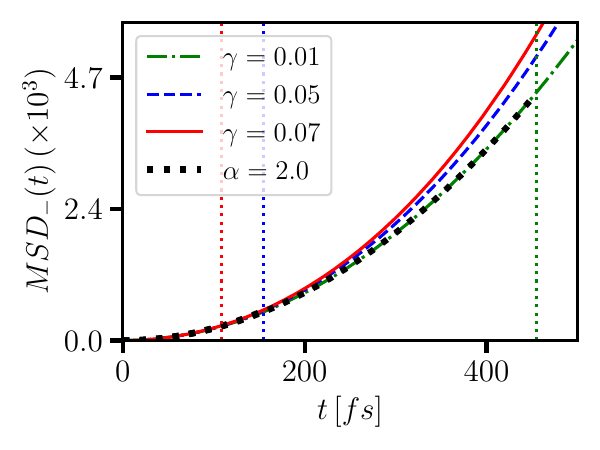}  
        \caption{}
        \label{fig:msd_lp_fit}
    \end{subfigure}
    \caption{(a) The population dynamics of the LP branch are shown for three photon loss rates: \(\gamma = 0.01\), \(0.05\), and \(0.07\). (b) The normalized center-of-mass (CM) position propagation of the LP branch is shown for the same set of photon loss rates. (c) Displays the mean squared displacement (MSD) of the LP wave packet dynamics. (d) The initial expansion of the wave packet exhibits ballistic transport, characterized by a power-law scaling of the form \( f_B t^2 \). A vertical line in the figure indicates this ballistic regime. Parameters: $\omega_M = 1.0, \omega_C = 0.4, g = 0.3, p = 0.03$}
    \label{fig:lower_polariton}
\end{figure}
%%%%%%%%%%%%%%%%%%%%%%%%%%%%%%%%%%%%%%%%%%%%%%%%%%%%%%%%%%%%%%%%%%%%%%%
We now turn to the LP branch, where the relaxation rate of the exciton populations increases linearly with the photon loss rate shown in Fig. \ref{fig:pop_lp_exc}. This behavior of the UP/LP branches are consistent with the prediction shown in Fig.~\ref{fig:decay_rate_g}(I).
%%%%%%%%%%%%%%%%%%%%%%%%%%%%%%%%%%%%%%%%%%%%%%%%%%%%%%%%%%%%%%%%%%%%%%%%%
\subsection{CM motion}
The motion of the center of mass (CM) of the polaritonic wave packet will be analyzed across different regimes of the photon loss rate. \textcolor{black}{In the continuum limit, we replace discrete sums with integrals, yielding the first moment as follows:
\begin{equation}
\braket{n(t)}\!=\!\sum_{n}nP_{n}(t)\!=\!-i\int\psi_{k}(t)\left.\frac{d\psi_{k+q}^{*}(t)}{dq}\right|_{q=0}dk.\label{eq:<n>-definition}
\end{equation}
The normalized first moment of the UP/LP populations, $\overline{\braket{n(t)}}_{\pm} = \braket{n(t)}_{\pm} / P^{M}_{\pm}$,  can be derived approximately from Eq. (\ref{eq:<n>-definition}) by retaining only the leading dynamical term (see Supplementary Section (IV))  for the full expression)}. The analytical form of $\overline{\braket{n(t)}}_{\pm}$ is
\begin{align}
\overline{\braket{n(t)}}_\pm & \approx \frac{t\int P_{k}(0)|\e_{k_{\pm}}|^{2} \partial_k \varepsilon_k{}_\pm e^{-\gamma_k{}_{\pm}t}\,dk}{\int P_{k}(0)|\e_{k_{\pm}}|^{2} e^{-\gamma_k{}_{\pm}t}\,dk},\label{eq:approximate-CM} 
\end{align}
where $\text{Re}[\partial \varepsilon_{k_\pm}] = v_g{}_{\pm}(k)$ is the group velocity.
The time derivative of Eq.~(\ref{eq:approximate-CM}) is written as
\begin{align}
\frac{d\overline{\braket{n(t)}}_\pm}{dt} & = \braket{v_g{}_{\pm}(k)}_{t} - t  \braket{\gamma_{k_\pm}v_g{}_{\pm}(k)}_{t} + t \braket{\gamma_k{}_{\pm}}_{t} \braket{v_g{}_{\pm}(k)}_{t}, \label{eq:derivative_cm}
\end{align}
where the first term $\braket{v_g{}_{\pm}(k)}_t$ is instantaneous average velocity and the second term $ \braket{\gamma_{k_\pm}v_g{}_{\pm}(k)}_{t} - \braket{\gamma_k{}_{\pm}}_{t} \braket{v_g{}_{\pm}(k)}_{t} $ is the correlation function between decay rate and group velocity. The CM motion of the UP is illustrated for different photon loss rates \((\gamma < \gamma_C,\, \gamma = \gamma_C,\, \gamma > \gamma_C)\) in Fig.~\ref{fig:cm_up}. The normalized CM of the UP, $\overline{\langle n(t) \rangle}_{+}$, exhibits a linear dependence on time, governed by the group velocity. In the regime \(\gamma < \gamma_C\), the wave packet propagates with a constant velocity in the forward direction. At the critical point \(\gamma = \gamma_C\), the propagation is nearly suppressed. For \(\gamma > \gamma_C\), the wave packet propagates in the backward direction, despite the positive sign of the initial momentum \(p\). This behavior reflects the nontrivial dependence of the group velocity on the photon loss rate, as illustrated in Fig.~\ref{fig:vg_g}(I).
  
Next, the LP dynamics of $\overline{\braket{n(t)}}_{-}$, shown in Fig.~\ref{fig:cm_lp}, are discussed for photon loss rates $\gamma = 0.01$, $0.05$, and $0.07$, and can be explained using Eq.~(\ref{eq:derivative_cm}).
For $\gamma = 0.01$, the dynamics of $\overline{\langle n(t)\rangle}_{-}$ exhibit linear growth, as the instantaneous group velocity dominates the behavior - a characteristic observed in lossless cavity systems. In the coherent regime with photon loss rate $\gamma = 0.05$, the dynamics exhibit three distinct temporal phases: (1). Early times $(t << 1/\gamma_k{}_{-})$: The system behavior is linear, dominated by the initial momentum where the first term of (\ref{eq:derivative_cm}) prevails as all \(k_-\) states contribute equally.
(2). Intermediate times: The dynamics become nonlinear as higher-$k$ states, with their enhanced group velocities, dominate the evolution. \textcolor{black}{This leads to accelerated motion where the first term and correlation correction terms surpass $t$ in Eq. (\ref{eq:derivative_cm})}.
(3). Late times $(t >> 1/\gamma_k{}_{-})$: All long-lived states eventually decay, with the explicit time dependence of the dissipative terms ensuring complete contraction of the wavepacket. By increasing the photon loss rate to $\gamma = 0.07$, the contraction occurs at an earlier time, as the time scale of contraction $t \propto 1/\gamma$ (detailed in SI (II.2)). In SI Fig.~IV.4 presents the dynamics of \( \overline{\langle n(t) \rangle}_{\pm} \) for a fixed photon loss rate \( \gamma < \gamma_C \), examined across three characteristic momenta \( p \) (i.e., \( p < k_r \), \( p = k_r \), and \( p > k_r \)). Notably, center-of-mass (CM) contraction is observed even for the UP branch when \( p > 0.92 \) (beyond resonance). The complete derivation and a detailed analysis of the CM dynamics are provided in SI Section IV.
%%%%%%%%%%%%%%%%%%%%%%%%%%%%%%%%%%%%%%%%%%%%%%%%%%%%%%%%%
\subsection{Ballistic to diffusive transition: MSD}
The dynamical spreading of a polaritonic wave packet (WP) can be characterized by its mean squared displacement (MSD).  
% \begin{align}
% \braket{n^{2}(t)} & =\sum_{n}n^{2}P_{n}(t) =-\int\psi_{k}(t)\left.\frac{d^{2}\psi_{k+q}^{*}(t)}{dq^{2}}\right|_{q=0}dk. \label{eq:<n^2>-definition}
% \end{align}
The MSD of wave packet \cite{tutunikov2025} is defined as
\begin{align}
\text{MSD}_{\pm}(t) =  \overline{\langle n^2(t)  \rangle}_\pm  -  \overline{\langle n(t) \rangle}^2_\pm - \frac{W^2}{2}, \label{eqn:msd}
\end{align}
where, \( W \) denotes the effective initial wave packet width, $\overline{\braket{n^{2}(t)}}_{\pm} = \braket{n^{2}(t)}/P^{M}_{\pm}$ and $P^{M}_{\pm}$ denotes the population in the molecular subsystem. \textcolor{black}{In the continuum limit, the second moment $\braket{n^{2}(t)}$ can be derived following the same methodology used for $\braket{n(t)}$ in Eq.~\ref{eq:<n>-definition} (see the complete analytical derivation in Supplementary Material~(IV). }

To analyze the ballistic to diffusive transition of the UP wave packet, we employ a fitting function.  
\begin{eqnarray}
f(t) = \alpha^2 \tau^2 \big( \frac{t}{\tau} - 1 + e^{-\frac{t}{\tau}} \big).\label{eqn:fitting_function}
\end{eqnarray}
The fitting function introduces two free parameters, $\alpha$ and $\tau$, which show excellent agreement with MSD in Eq.~\ref{eqn:msd}. As illustrated in Fig.~\ref{fig:msd_up}, these parameters capture the transition between coherent and incoherent regimes through the critical point. In the coherent regime $(\gamma < \gamma_{C})$, the wave packet initially propagates ballistically before transitioning to diffusion at long times. At the critical point $\gamma = \gamma_C$, the dynamics become purely diffusive. Interestingly, for $\gamma > \gamma_C$, the ballistic regime re-emerges. From our fitting function, we extract the ballistic-to-diffusive transition time $\tau$, which is plotted as a function of $\gamma$ in Fig.~\ref{fig:transition_gamma}. When $\gamma$ attains its critical value $\gamma_C$, the ballistic propagation is fully suppressed the ballistic propagation, leaving only diffusive motion. The detailed analysis of the fitting function and its comparison with the MSD scaling law, \( \text{MSD}(t) \propto t^{\beta} \), where the exponent $\beta$ characterizes the transport properties, are presented in SI~(V.1). Additionally, we discuss the MSD dynamics for both resonance and off-resonance cases ( \(i.e., k \) dependent) in SI~(V.2).

The MSD dynamics of the LP branch, shown in Fig.~\ref{fig:msd_lp}, exhibit three distinct temporal phases: (1) ballistic, (2) nonlinear overshoot, and (3) contraction, similar to the contraction observed in the velocity dynamics in Fig.~\ref{fig:cm_lp}. The ballistic phase shows excellent agreement with the scaling law \( \text{MSD}(t) \propto t^2 \), followed by rapid expansion in the intermediate (nonlinear) regime and subsequent contraction. This behavior can be experimentally verified using the technique described in \cite{pandaya2022sciadv}. As the photon loss rate increases, the dynamics decay more rapidly at early times, and the ballistic range decreases, becoming evident on a timescale of $\gamma^{-1}$.

%%%%%%%%%%%%%%%%%%%%%%%%%%%%%%%%%%%%%%%%%%%%%%%%%%%%%%%%%%%%%%%%%%%%%%%%%%%%
In addition to the experimental observation of UP branch contraction reported in Ref.~\cite{pandaya2022sciadv}, our results reveal contraction dynamics in the LP branch for \( k < k_r \) (before resonance) and in the UP branch for \( k > k_r \) (beyond resonance). The detailed expression of MSD and analysis for both branches are provided in SI~(V). %\ref{si:msd_fitting_function}.
%%%%%%%%%%%%%%%%%%%%%%%%%%%%%%%%%%%%%%%%%%%%%%%%%%%%%%%%%%%%%%%%%%%%%%%%%%%%%%%%%%%%%%%%%%%%%%%
\section{Conclusions}
In this work, we present a systematic investigation of the multimode non-Hermitian Tavis-Cummings (MNTC) model. In the first half of the paper, we analyze the complex eigen-structure, including the real part of the eigenvalue (i.e., dispersion), $\text{Re}[\varepsilon_{k_\pm}]$, its derivative (i.e., the group velocity) $v_g = \text{Re}[\varepsilon'_{k_\pm}] $), and the imaginary part of the eigenvalue (i.e., relaxation rate), $\text{Im}[\varepsilon_{k_\pm}]$, as functions of the wave vector $k$ for three characteristic photon decay rates $\gamma$. These photon decay rates correspond to the \textit{coherent}, \textit{critical}, and \textit{incoherent} dynamical regimes. In addition, we investigate the dependences of these quantities on the photon loss rate $\gamma$ for three representative wave vectors, representing below-resonance, on-resonance, and above-resonance. Our key observations are summarized as follows:

\begin{itemize}
\item Dispersion:  As a function of wavevector \(k \), 
$\text{Re}[\varepsilon_{k_\pm}]$ of the UP/LP branches shows avoided crossing below the resonance $( k < k_r)$, exact crossing at resonance $( k = k_r)$, and flattening crossing above the resonance $( k > k_r)$. As a function of the photon loss rate $\gamma$,  at off-resonance, the eigen energy shows weak $\gamma$-dependence. Remarkably, at resonance ($k=k_r$), the Rabi splitting decreases with $\gamma$ until eigen energies coalesce, resembling the exceptional point behavior in disordered systems\cite{engelhardt2022prb} shown in Figs.~\ref{fig:dispersion_k} and \ref{fig:dispersion_g}.

\item Group Velocity: As a function of wavevector \( k \), at the coherent rate ($\gamma < \gamma_C$), the UP branch increases monotonically and the LP branch peaks. The two curves crosses at resonance. The transition to the incoherent regime ($\gamma \geq \gamma_C$) introduces anomalous transport, manifested through negative group velocities. The $\gamma$-dependence shows the critical behavior at $\gamma_C$, and the sign reversal in $v_g$ depends sensitively on the resonance condition, demonstrating the interplay between non-Hermitian dynamics and light-matter hybridization shown in Figs~\ref{fig:vg_k} and \ref{fig:vg_g}.  

\item Relaxation Rate:  As a function of wavevector (\(k \)), below the critical point, $\gamma < \gamma_c$, $\text{Im}[\varepsilon_{k_\pm}]$ increases monotonically in the UP branch initially and decreases monotonically in the LP branch, with crossing at resonance.  Beyond the critical point, a gap opens between the UP and LP branches, marking the coherent-incoherent transition. 
As a function of the photon loss rate $\gamma$,  below resonance ($k<k_r$), the UP branch exhibits a turnover in the $\gamma$-dependence with a maximum at $\gamma_c$~\cite{andrew2025nano}, while the LP branches scale linearly. The roles of the two branches are reversed above the resonance ($k>k_r$).  At resonance ($k=k_r$), a single relaxation rate in the coherent regime bifurcates at $\gamma_c$ to two parabolic branches in the incoherent regime, as shown in Figs.~\ref{fig:decay_rate_k} and \ref{fig:decay_rate_g}.

\end{itemize}

The above theoretical analysis leads to distinct wave packet dynamics of the UP/LP branches in the MNTC model. In the second half of the paper, we report wave-packet propagation for three values of the photon loss rates below resonance and observe striking effects of photon loss rate $\gamma$ on population relaxation, CM motion, and MSD.

\begin{enumerate}
\item{ Population Relaxation: The UP branch shows a non-monotonic relaxation rate which peaks at the critical point $\gamma_c$, while the relaxation rate of the LP branch grows linearly with the photon loss rate, consistent with the spectral analysis. }
\item{ CM Motion: The UP branch exhibits a reversal in the direction of its center-of-mass motion when the photon loss rate exceeds its critical point $\gamma>\gamma_c$, marking the coherent-to-incoherent transition. In contrast, the LP branch exhibits three dynamical phases: an initial ballistic propagation, an accelerated phase, and a subsequent spatial contraction, in agreement with recent simulations~\cite{tichauer2023sca}.}
\item{ MSD Dynamics:  The UP branch exhibits a transition from ballistic to diffusive motion, which has been observed in recent experimental observations and numerical simulations and is often attributed to static and dynamic disorder.~\cite{mukundakumar2023, xu2023natcomm}. 
The transition time also shows a crossover as a function of the photon loss rate and vanishes at the critical point, indicating purely diffusive motions. In contrast, the LP branch displays a distinct three-phase behavior characterized by the initial ballistic motion, nonlinear overshoot, and eventual contraction, in agreement with recent experimental findings~\cite{pandaya2022sciadv}.}
\end{enumerate}

We performed an extensive study over a broad range of momenta spanning the below-resonance, resonance, and above-resonance regimes. Owing to the symmetry around the resonance point, similar dynamical features emerge beyond resonance in the opposite polariton branch. Thus, the phase diagram shows complementary symmetry with respect to the resonance and coherent-to-incoherent transition.

\bibliography{main}

\end{document}